\documentclass[a4paper]{jpconf}

\usepackage{graphicx}

\newcommand{\Fi}[1]   {Fig.~\ref{#1}}

\newcommand{\agev}    {\mbox{$A$~GeV}}               

\newcommand{\rb}[1]   {\mbox{\textrm{\scriptsize #1}}}
\newcommand{\rbt}[1]  {\mbox{\textrm{\tiny #1}}}
\newcommand{\sqrts}   {\ensuremath{\sqrt{s_{_{\rbt{NN}}}}}}

\newcommand{\lam}     {\ensuremath{\Lambda}}
\newcommand{\lab}     {\ensuremath{\bar{\Lambda}}}  
\newcommand{\xim}     {\ensuremath{\Xi^{-}}}
\newcommand{\xizero}  {\ensuremath{\Xi^{0}}}
\newcommand{\xip}     {\ensuremath{\bar{\Xi}^{+}}}

\newcommand{\pim}     {\ensuremath{\pi^-}}
\newcommand{\pip}     {\ensuremath{\pi^+}}

\newcommand{\pbar}    {\ensuremath{\bar{\textrm{p}}}}

\newcommand{\mt}      {\ensuremath{m_{\rb{t}}}}
\newcommand{\mtmzero} {\ensuremath{m_{\rb{t}}-m_{\rb{0}}}}
\newcommand{\mtavg}   {\ensuremath{\langle m_{\rb{t}} \rangle - m_{\rb{0}}}}

\newcommand{\dndy}    {\ensuremath{\textrm{d}n/\textrm{d}y}}

\newcommand{\npart}   {\ensuremath{N_{\rb{part}}}}

\newcommand{\mub}     {\ensuremath{\mu_{\rbt{B}}}}
\newcommand{\gams}    {\ensuremath{\gamma_{\rb{s}}}}

\newcommand{\ybeam}   {\ensuremath{y_{\rb{beam}}}}

\begin{document}


\title{Particle Production at the SPS and the QCD Phase Diagram}

\author{Christoph Blume}

\address{Institut f\"ur Kernphysik, J.W.~Goethe-Universit\"at,  \\
Max-von-Laue-Str.~1, D-60438 Frankfurt am Main, GERMANY }

\ead{blume@ikf.uni-frankfurt.de}


\begin{abstract}
Recent results of particle production in the energy regime of the
CERN-SPS are reviewed.  In order to collect information on the 
properties of the QCD phase diagram systematic studies of the system 
size and the energy dependence of particle production in heavy ion 
collisions have been performed.  Net-baryon distributions and results
on strangeness production are discussed.  The system size dependence of 
many observables can be understood in the core-corona approach, which
has implications on the possibility to use system size as a control
parameter to study different areas of the phase diagram.  Recent
attempts to search for a critical point, such as multiplicity
fluctuations and the transverse mass dependence of anti-baryon/baryon
ratios are reviewed.
\end{abstract}


\section{Introduction}
\vspace{16pt}

The QCD phase diagram, as it is illustrated in \Fi{fig:phasediag}, contains 
a variety of theoretically predicted features.  The most important one
is the phase boundary separating a hadron gas and a quark gluon plasma
(red dashed line in \Fi{fig:phasediag}).  Its position at vanishing baryonic 
chemical potential ($\mub = 0$) is defined by the critical temperature 
$T_{\rb{C}}$.  Even though recent attempts to calculate the exact value of 
$T_{\rb{C}}$ with lattice QCD still result in different numerical values 
\cite{AOKI1,CHENG1,FODOR2}, there is a general consensus that for $\mub = 0$ 
this transition is of the type of crossover.  However, this crossover line 
might turn for a certain \mub\ into a first order phase transition, which 
would result in a critical point at that position \cite{STEPHANOV3}.  
However, predictions on the location of the critical point are still very
uncertain \cite{STEPHANOV4}.  It might very well be that it does not exist
at all \cite{FORCRAND1}.  Recently, it has been conjectured that the phase 
diagram might contain yet another phase, the so-called quarkyonic phase 
\cite{MCLERRAN1}, which would be separated from normal hadronic matter by 
a different phase boundary line (blue dashed line in \Fi{fig:phasediag}).

Experimentally, there are several control parameters available that might
allow to study different regions of the phase diagram with heavy ion
collisions.  One is the variation of the center-of-mass energy which
will force the reaction systems to follow different trajectories in the
$T$-\mub~plane, reflected by a change of the chemical freeze-out parameters.
The second control parameter is the variation of the system size, which
can either be achieved by performing central collisions of nuclei of 
different size, or by studying centrality selected minimum bias collisions.
However, whether this can be considered as a good control parameter for 
probing different areas of the phase diagram shall be discussed further
below.

%
\begin{figure}[th]
\begin{center}
\includegraphics[width=0.5\linewidth]{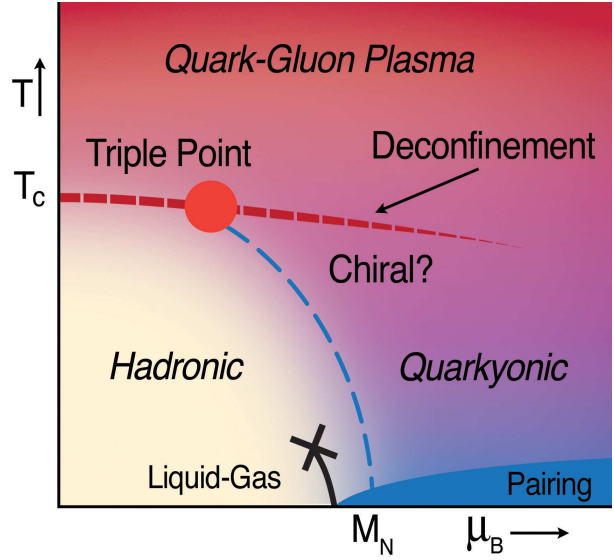}
\end{center}
\caption{The phase diagram of strongly interacting matter.  The picture 
is taken from \cite{ANDRONIC1}.
}
\label{fig:phasediag}
\end{figure} 
%

In the following we review some results from systematic studies of 
particle production in heavy ion collisions performed at the CERN-SPS,
and will discuss what can be learned from existing data on the experimental 
accessibility of different areas in the phase diagram.  Finally, we also 
present some results from existing attempts to search for a critical point.


\section{Variation of Energy}
\vspace{16pt}

Recent results on rapidity spectra of (anti-)protons in central Pb+Pb
reactions, measured at energies between 20$A$ - 158\agev\ at the SPS 
\cite{NA49STOP,NA49EDEPPR}, together with central Au+Au data from the AGS 
\cite{E802PROT} and RHIC \cite{BRMSSTOP} allow to study the energy evolution 
of stopping.  Based on the measured rapidity spectra for p, $\bar{\textrm{p}}$, 
\lam, \lab, \xim, and \xip, all corrected for feed down from weak decays, the 
net-baryon distributions $\bar{\textrm{B}} - \textrm{B}$ have been 
constructed \cite{BLUME1}.  The contribution of unmeasured baryons (n, 
$\Sigma^{\pm}$, \xizero) was estimated using the results of a statistical 
hadron gas model \cite{BECATTINI3}.  In the SPS energy region a clear 
evolution of the shape can be observed (see left panel of 
\Fi{fig:edep_baryons}), changing from a single maximum at midrapidity 
towards a structure with two maxima and a dip at $y = 0$.  Concurring with 
this evolution, also a strong change of the anti-baryon/baryon ratios takes 
place in the energy region of the SPS \cite{NA49EDEPPR,NA49EDEPHYP} and thus 
of the baryonic chemical potential at freeze-out \mub.

%
\begin{figure}[th]
\begin{center}
\begin{minipage}[b]{0.43\linewidth}
\begin{center}
\includegraphics[width=\linewidth]{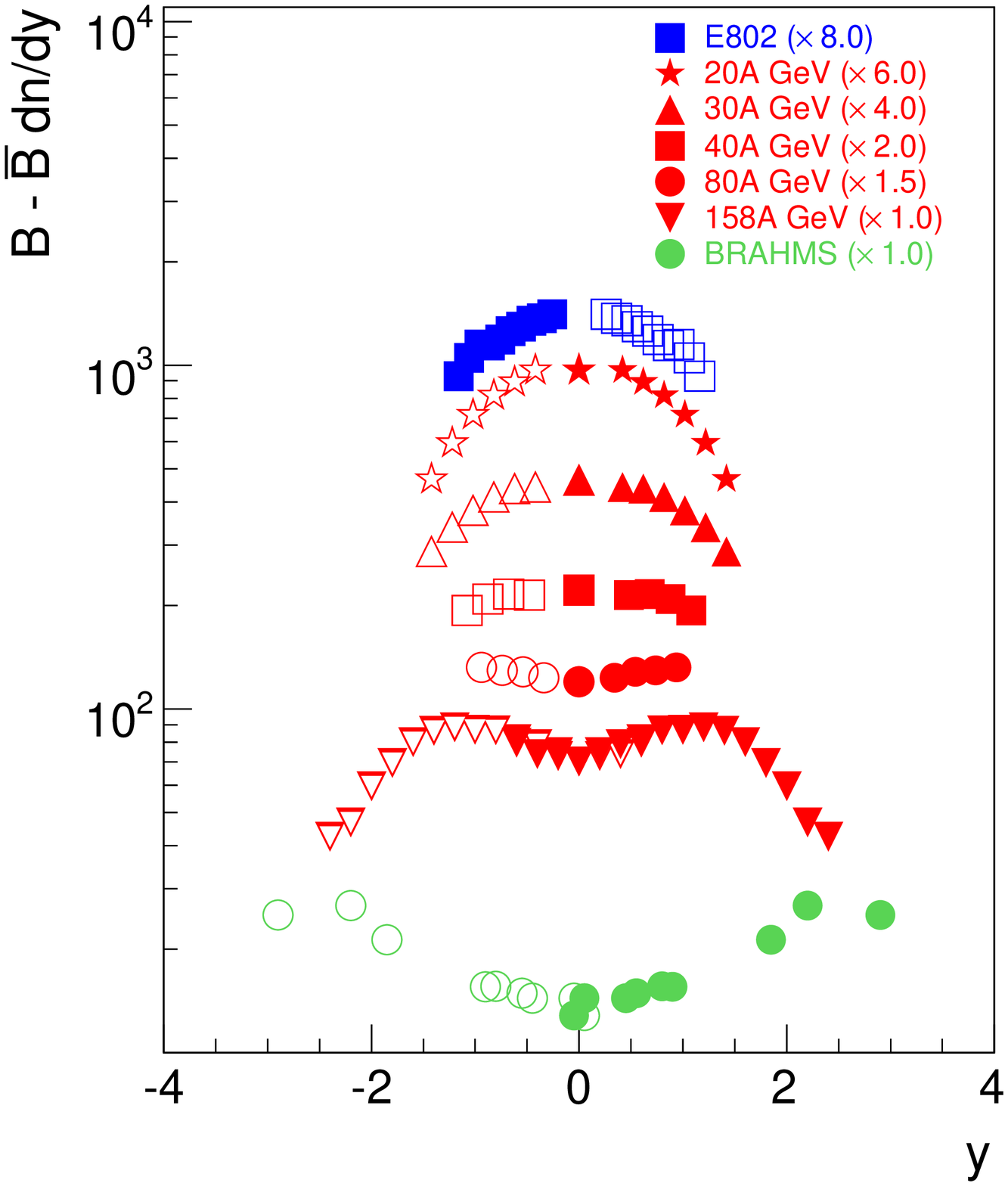}
\end{center}
\end{minipage}
\begin{minipage}[b]{0.55\linewidth}
\begin{center}
\includegraphics[width=\linewidth]{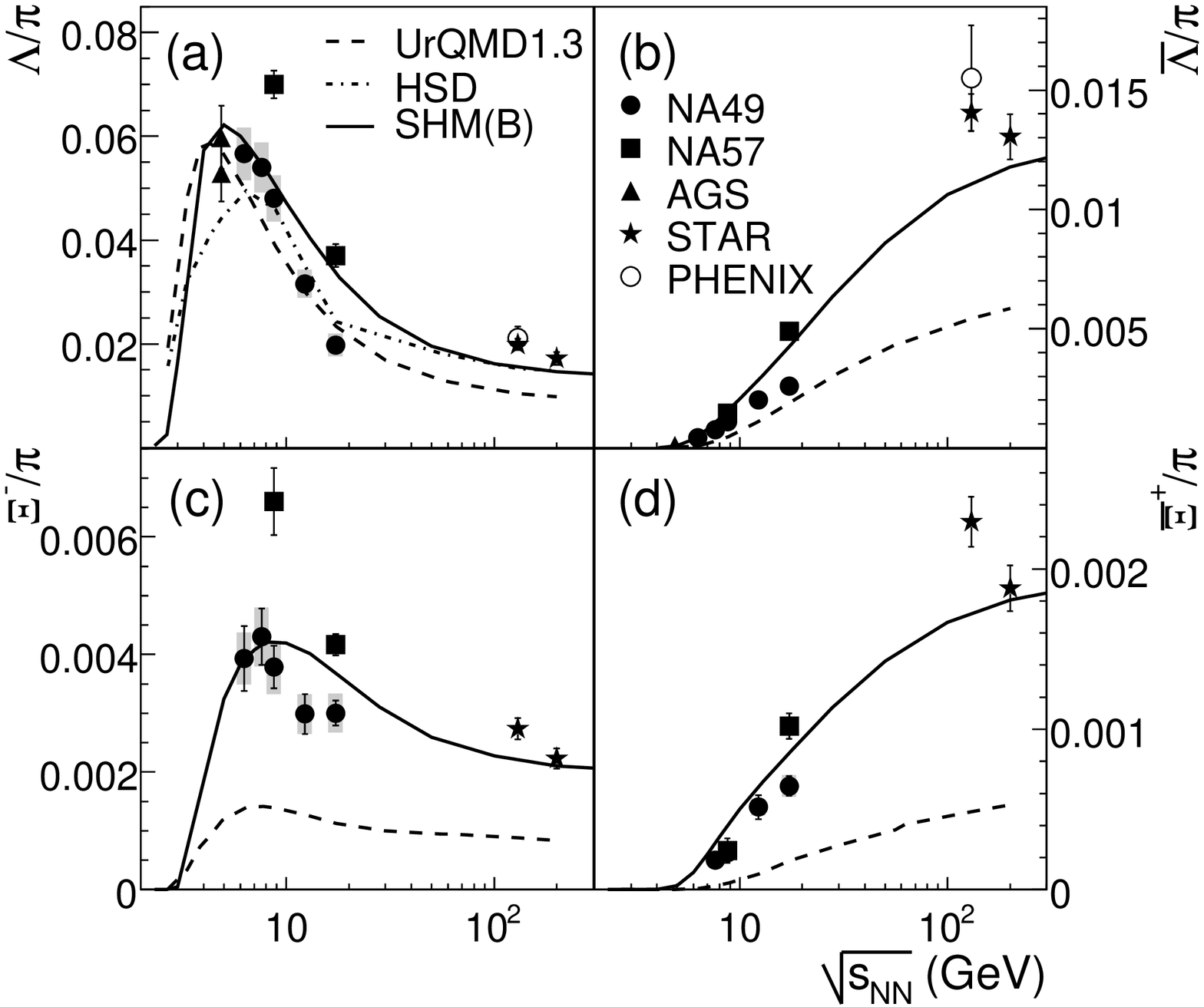}
\end{center}
\end{minipage}
\end{center}
\caption{Left: The rapidity distributions of net-baryons at SPS
energies~\cite{BLUME1,NA49STOP,NA49EDEPPR}, together with results from the 
AGS~\cite{E802PROT} and from RHIC~\cite{BRMSSTOP} for central 
Pb+Pb(Au+Au) collisions.
Right: The energy dependence of the rapidity densities \dndy\ around 
mid-rapidity of \lam, \lab, \xim, and \xip, divided by the total pion rapidity 
densities (\mbox{$\pi = 1.5 \:(\pim + \pip)$}) for central Pb+Pb and Au+Au
collisions \cite{E896LAM,E917LAB,E891LAM,E802PI116,NA49EDEPHYP,NA57ENHANCE,
NA57EDEPHYP,STARLAM130,STARHYP130,PHNXLAM130,STARHYP200,STARKPI130,
STARKPM200,PHNXKPI130}.  Also shown are results for the transport models
UrQMD1.3 (dashed line) \cite{URQMD1} and HSD (dotted line) \cite{HSD}, as well 
as a statistical hadron gas model (solid line) \cite{ANDRONIC2}.
}
\label{fig:edep_baryons}
\end{figure} 
%

The right panel of \Fi{fig:edep_baryons} shows a comparison of the energy 
dependence of mid-rapidity \lam, \lab, \xim, and \xip\ production to 
several models.  While the transport models UrQMD1.3 \cite{URQMD1} and HSD
\cite{HSD} provide a reasonable description of the \lam/$\pi$ and \lab/$\pi$ 
ratios, they are clearly below the data points in case of the \xim\ and \xip.  
This might indicate that an additional partonic contribution is necessary to 
reach the production rates observed for multi-strange particles.  Statistical 
models on the other hand, generally provide a better match to the data.  These 
models are based on the assumption that the particle yields correspond to 
their chemical equilibrium value and can thus be described by the parameters 
temperature $T$, baryonic chemical potential \mub, volume $V$, and, in some 
implementations, by an additional strangeness under-saturation factor \gams. 
The curves shown in the right panel of \Fi{fig:edep_baryons}, labeled SHM(B), 
are taken from \cite{ANDRONIC2} and are based on parametrizations of the 
\sqrts\ dependence of $T$ and \mub, assuming \gams~=~1.  The results of 
various analysis of the measured particle abundances with statistical model 
approaches show that $T$ and \mub\ parameter for the SPS energy range vary 
over a wide region, following an universal freeze-out curve \cite{CLEYMANS2}, 
and might be in the vicinity of the possible critical point location 
\cite{ANDRONIC2,BECATTINI4,CLEYMANS1,FODOR1}.  Thus, the variation of \sqrts\ 
provides a well defined way of selecting different regions in the $T$-\mub~plane.


\section{Variation of System Size}
\vspace{16pt}

\subsection{Centrality Dependence of Net Protons}

\begin{figure}[th]
\begin{center}
\begin{minipage}[b]{0.49\linewidth}
\begin{center}
\includegraphics[width=\linewidth]{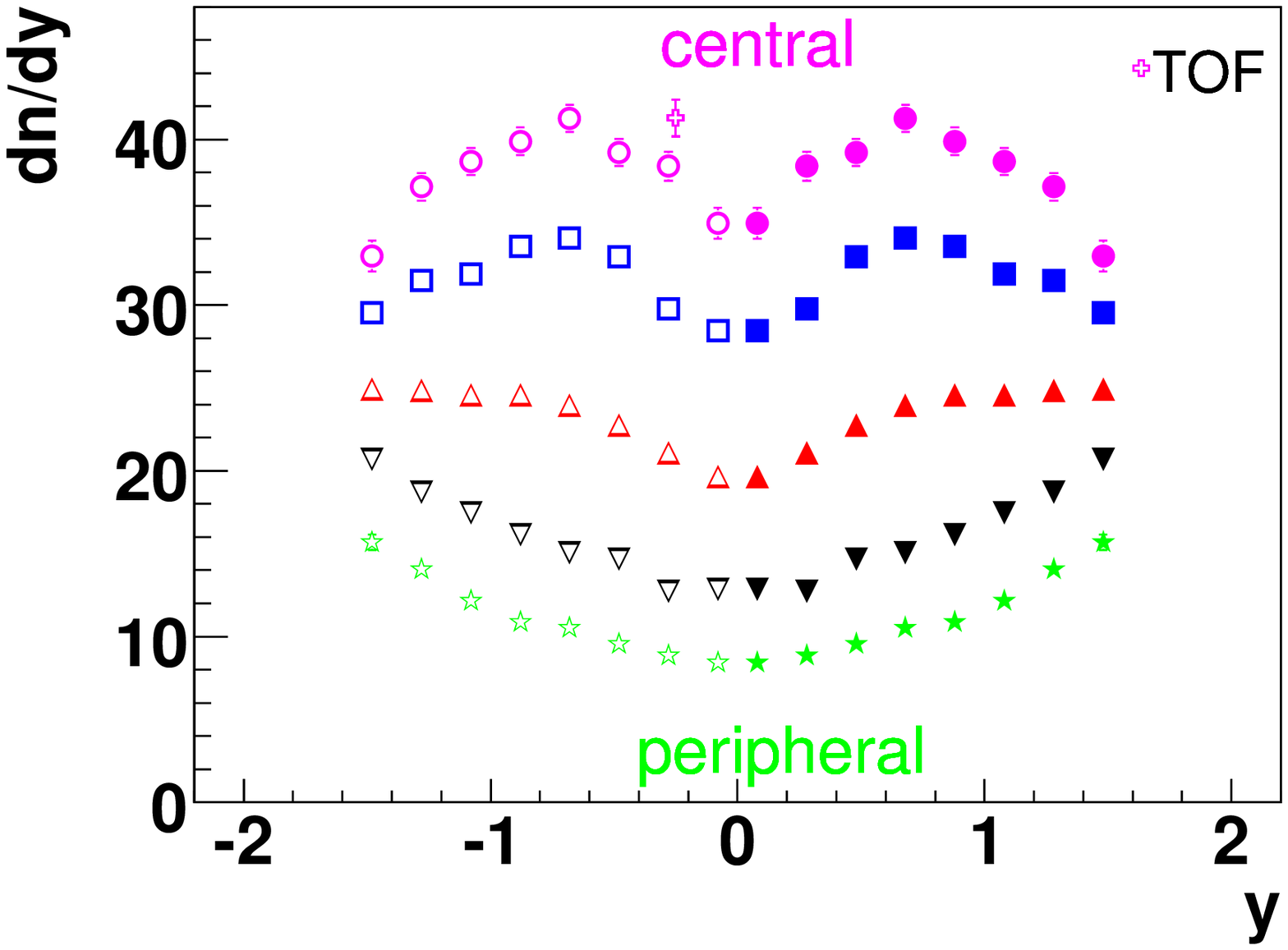}
\end{center}
\end{minipage}
\begin{minipage}[b]{0.49\linewidth}
\begin{center}
\includegraphics[width=\linewidth]{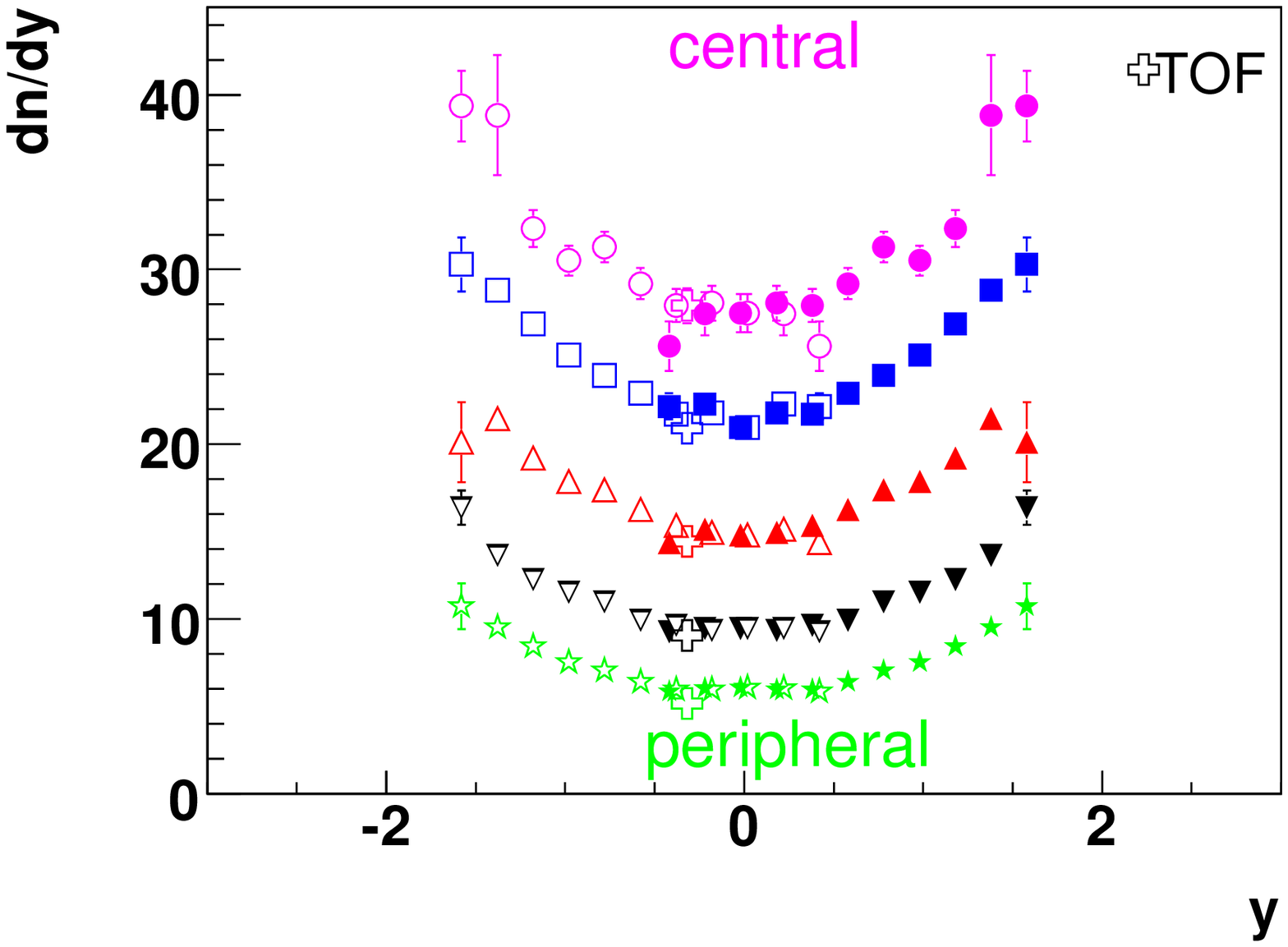}
\end{center}
\end{minipage}
\end{center}
\caption{Net-proton rapidity distributions for Pb+Pb collisions at 40$A$ 
(left) and 158\agev\ (right) for different centralities as measured by
the NA49 experiment \cite{STROEBELE}.  The full symbols represent the 
measurements.  The open symbols are reflected at midrapidity.
}
\label{fig:net_protons} 
\end{figure}

Recent data on proton and antiproton production in minimum bias Pb+Pb 
reactions allow to study the system size dependence of stopping. 
Figure~\ref{fig:net_protons} shows the net-proton rapidity distributions 
for five different centrality classes, selected from minimum bias Pb+Pb 
interactions at 40$A$ (left panel) and 158\agev\ (right panel) 
\cite{STROEBELE}.  A remarkable feature of this data is that there is 
no drastic change with centrality of the shapes of the distributions 
inside the measured region at 158\agev.  However, at 40\agev\ the shapes 
of the distributions change with centrality, going from a two-maxima
structure with a narrow dip at midrapidity for central collisions to a 
shallow valley like distribution for peripheral collisions, similar as
observed at the AGS \cite{E917PROT}.

%
%

The \pbar/p~ratio, being directly sensitive to the baryonic chemical 
potential \mub, depends much less on system size than it does depend
on energy.  While for the mid-rapidity ratios at 158\agev\ only a change
in the order of a factor of two is observed when comparing central Pb+Pb 
reactions (\pbar/p = 0.56, 0 -- 5\% most central) to very peripheral ones 
(\pbar/p = 0.11 for 43 -- 100\%), this ratio varies by a factor of 43 
between central Pb+Pb collisions at 20\agev\ (\pbar/p = 0.0013, 0 -- 7\%) 
and 158\agev\ \cite{NA49EDEPPR}.  Thus, only a small variation of \mub\ can 
be achieved when studying reactions of different system sizes.

\subsection{Core Corona Model}

It has been found that the core-corona approach is able to describe the 
system size dependence of particle production reasonable well
\cite{BOZEK,BECATTINI1,BECATTINI2,AICHELIN1,WERNER1}.  In this model a 
nucleus-nucleus collision is decomposed into a central core, which 
corresponds to the large fireball produced in central A+A collisions,
and a peripheral corona, that is equivalent to independent nucleon-nucleon
reactions.  To quantify the relative contribution of the two components,  
the fraction of nucleons that scatter more than once $f(\npart)$ can be
used. $f(\npart)$ can simply be calculated with a Glauber model \cite{GLAUBER}.  
This quantity allows for a natural interpolation between the yields $Y$ 
measured in elementary p+p ($= Y_{\textrm{corona}}$) and in central 
nucleus-nucleus collisions ($= Y_{\textrm{core}}$):
\begin{equation}
Y(\npart)  =  \npart \: [ f(\npart) \: Y_{\textrm{core}}  \: 
                        + \: (1 - f(\npart)) \: Y_{\textrm{corona}} ]
\end{equation}
The left panel of \Fi{fig:meanmt_vs_nw} shows for example the result of 
this approach compared to the system size dependence of the 
strangeness enhancement factors $E$ at 158\agev\ \cite{NA49HYSDEP,BLUME2}.
Here, $E$ is defined relative to p+p reactions as baseline measurement:
\begin{equation}
E = \left. \left( \frac{1}{\langle \npart \rangle} 
           \left. \frac{\textrm{d}N(\textrm{Pb+Pb})}{\textrm{d}y}\right|_{y=0}
    \right) \right/
    \left( \frac{1}{2} 
           \left. \frac{\textrm{d}N(\textrm{p+p})}  {\textrm{d}y}\right|_{y=0}
    \right)
\end{equation}
The observed rapid rise for very small systems and the subsequent saturation
can naturally be described within this model.

It is interesting to observe that this approach not only works for yields, 
but also for dynamical quantities such as the average transverse momenta 
\mtavg\ (see right panel of \Fi{fig:meanmt_vs_nw}).  This suggests that the 
core-corona approach in general provides a reasonable way for understanding 
the evolution from elementary p+p to central Pb+Pb collisions.

%
\begin{figure}[th]
\begin{center}
\begin{minipage}[b]{0.45\linewidth}
\begin{center}
\includegraphics[width=\linewidth]{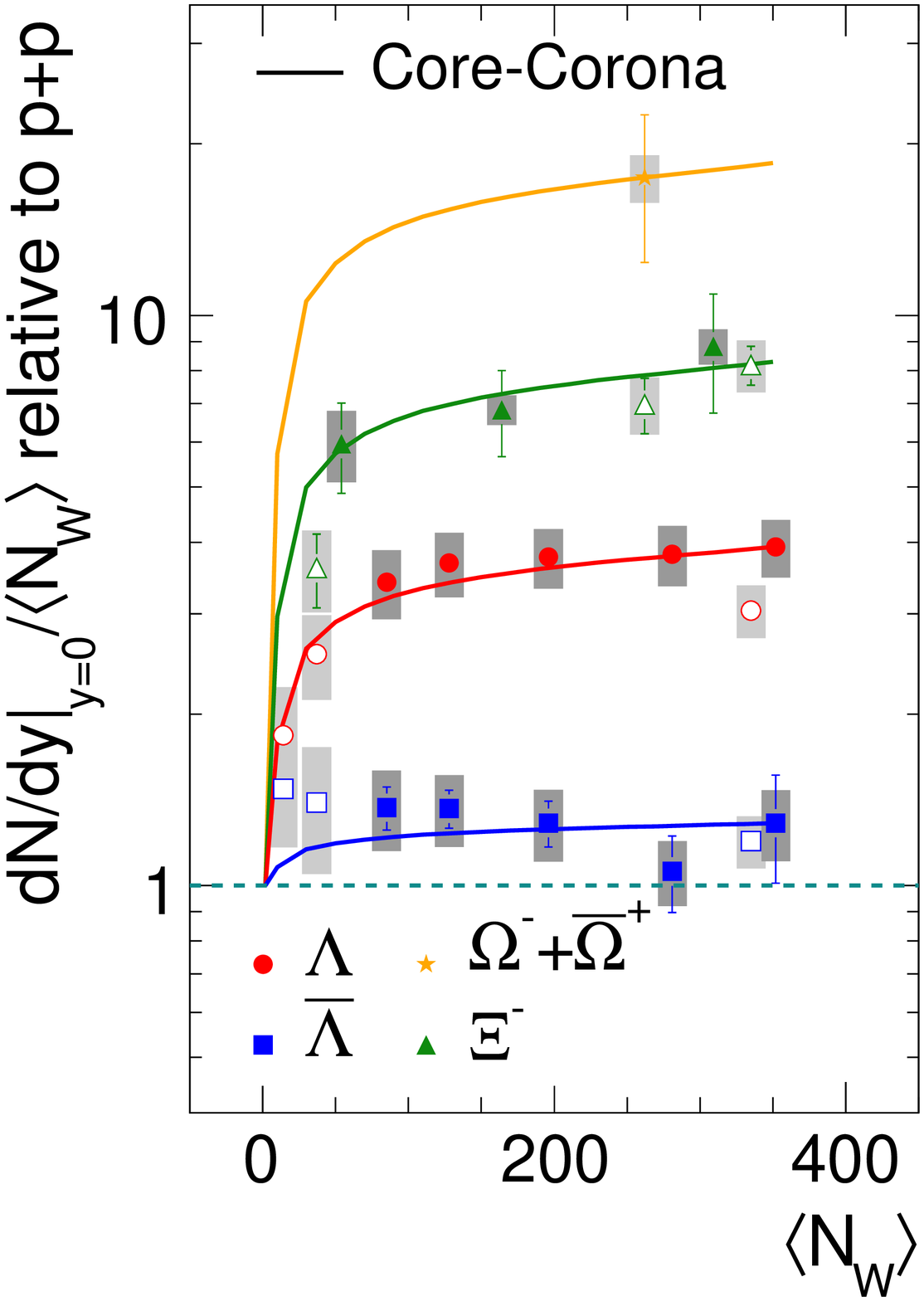}
\end{center}
\end{minipage}
\begin{minipage}[b]{0.45\linewidth}
\begin{center}
\includegraphics[width=\linewidth]{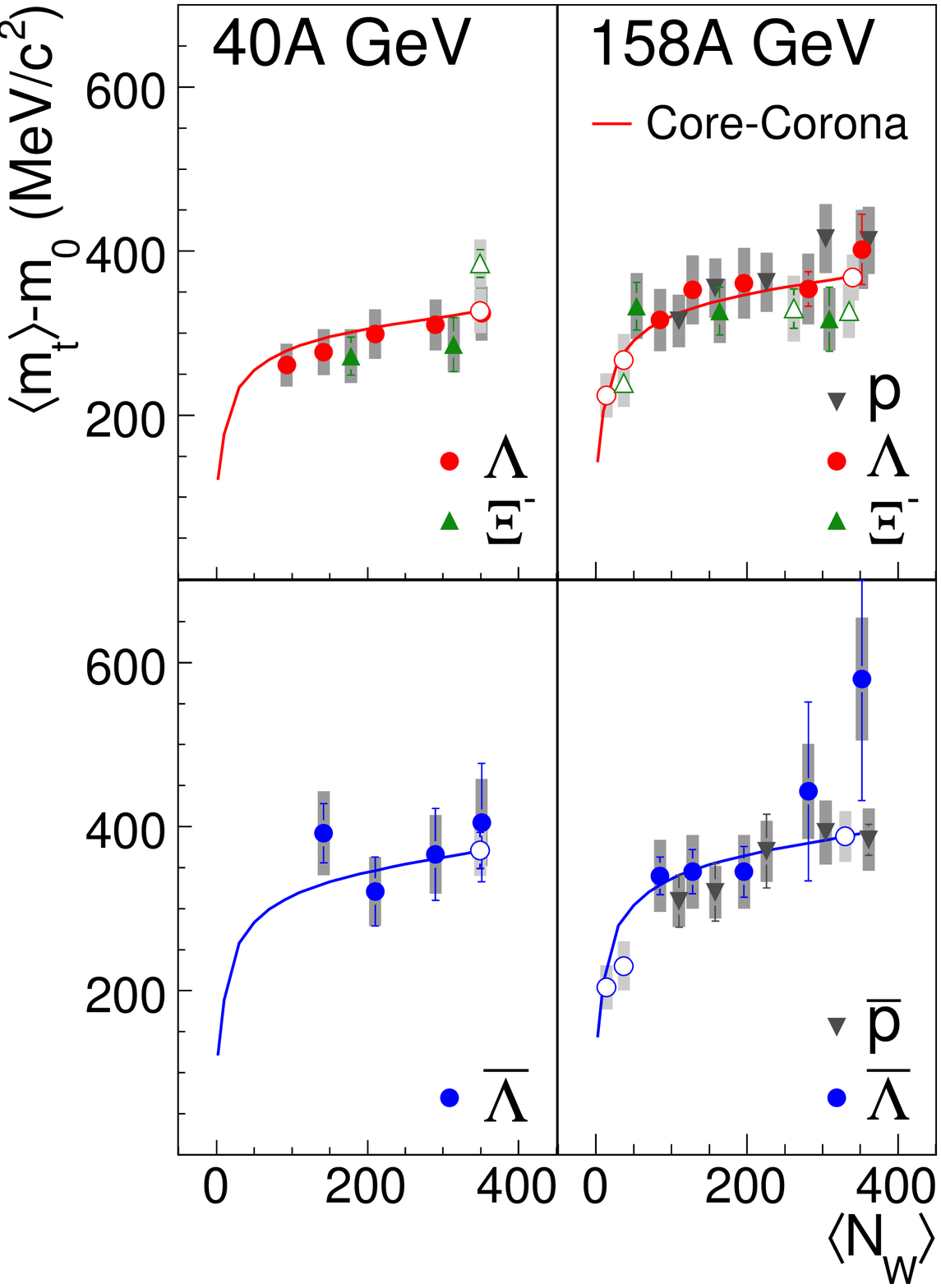}
\end{center}
\end{minipage}
\end{center}
\caption{
Left: The midrapidity yields per wounded nucleon relative to p+p yields
for central C+C, Si+Si and minimum bias Pb+Pb reactions at 158\agev\
\cite{NA49HYSDEP,BLUME2}.
Right: The \mtavg\ values at mid-rapidity for Pb+Pb collisions at 40$A$ 
and 158\agev, as well as for near-central C+C and Si+Si reactions at 
158\agev\ \cite{NA49HYSDEP}.  The (anti-)proton data are taken from 
\cite{NA49SDPR1}.  Also shown are the results from a fit for \lam\ and 
\lab\ with the core-corona approach (solid lines).
}
\label{fig:meanmt_vs_nw} 
\end{figure} 
%

A more differential look on the core-corona model is shown in the left
panel of \Fi{fig:glauber} for symmetric reaction systems.  Due to their 
different surface to volume ratio, the \npart\ dependence of $f(\npart)$ 
is much steeper for smaller reaction systems than for larger ones\footnote{
Please note that the curves shown in \Fi{fig:meanmt_vs_nw} are based on a 
function $f(\npart)$ that was calculated for Pb+Pb interactions 
\cite{AICHELIN1}.  Therefore their direct comparison to the smaller systems 
C+C and Si+Si is not possible.}  Also, the maximum value of $f(\npart)$ 
depends on the size of the nuclei.  E.g., while for very central Pb+Pb 
collisions $f_{\rb{max}}(\npart) \approx 0.9$ can be reached, the maximum 
value for very central C+C reactions is $f_{\rb{max}}(\npart) \approx 0.65$.  
This $A$ dependence of $f_{\rb{max}}$ is illustrated by the dashed red line 
in the left panel of \Fi{fig:glauber}.  This has the consequence that the
different relative core corona contributions have to be taken into account 
when comparing even very central nucleus-nucleus collisions of different 
size.  Especially, this might explain any apparent system size dependence 
of the chemical freeze-out parameters $T$ and \mub, as determined by 
statistical model fits to central A+A collisions of different size (e.g. 
\cite{BECATTINI3}).  Therefore, we conclude that, assuming the core-corona 
model is valid, the system size might not provide a good control parameter 
to probe different regions of the QCD phase diagram with produced fireballs 
of different temperature.  It is more likely that one only observes a change 
in the relative admixture of central fireball (core) and peripheral p+p like 
corona, whose freeze-out parameters are different, but independent of the 
size of the involved nuclei.

%
\begin{figure}[th]
\begin{center}
\begin{minipage}[b]{0.49\linewidth}
\begin{center}
\includegraphics[width=\linewidth]{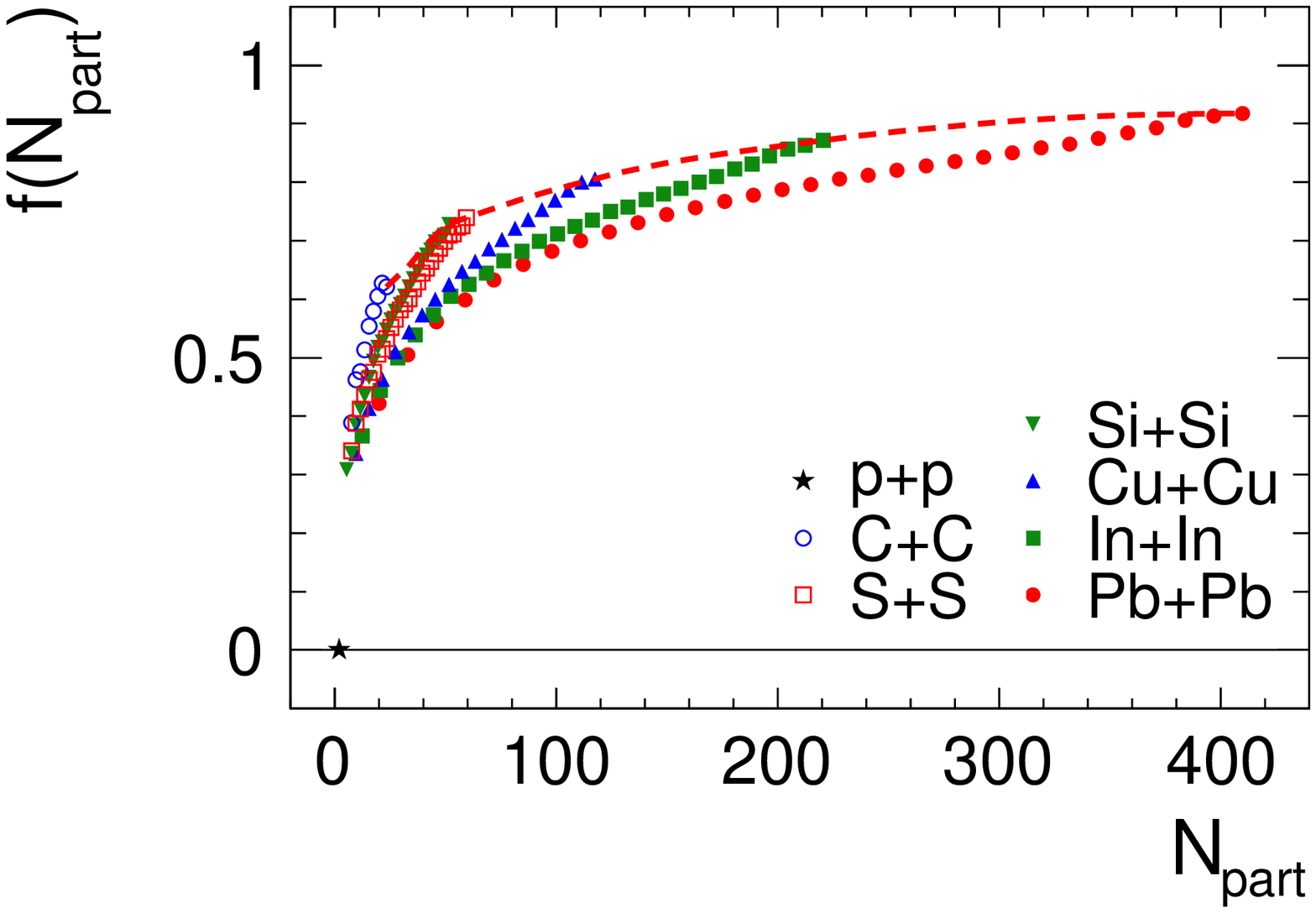}
\end{center}
\end{minipage}
\begin{minipage}[b]{0.49\linewidth}
\begin{center}
\includegraphics[width=\linewidth]{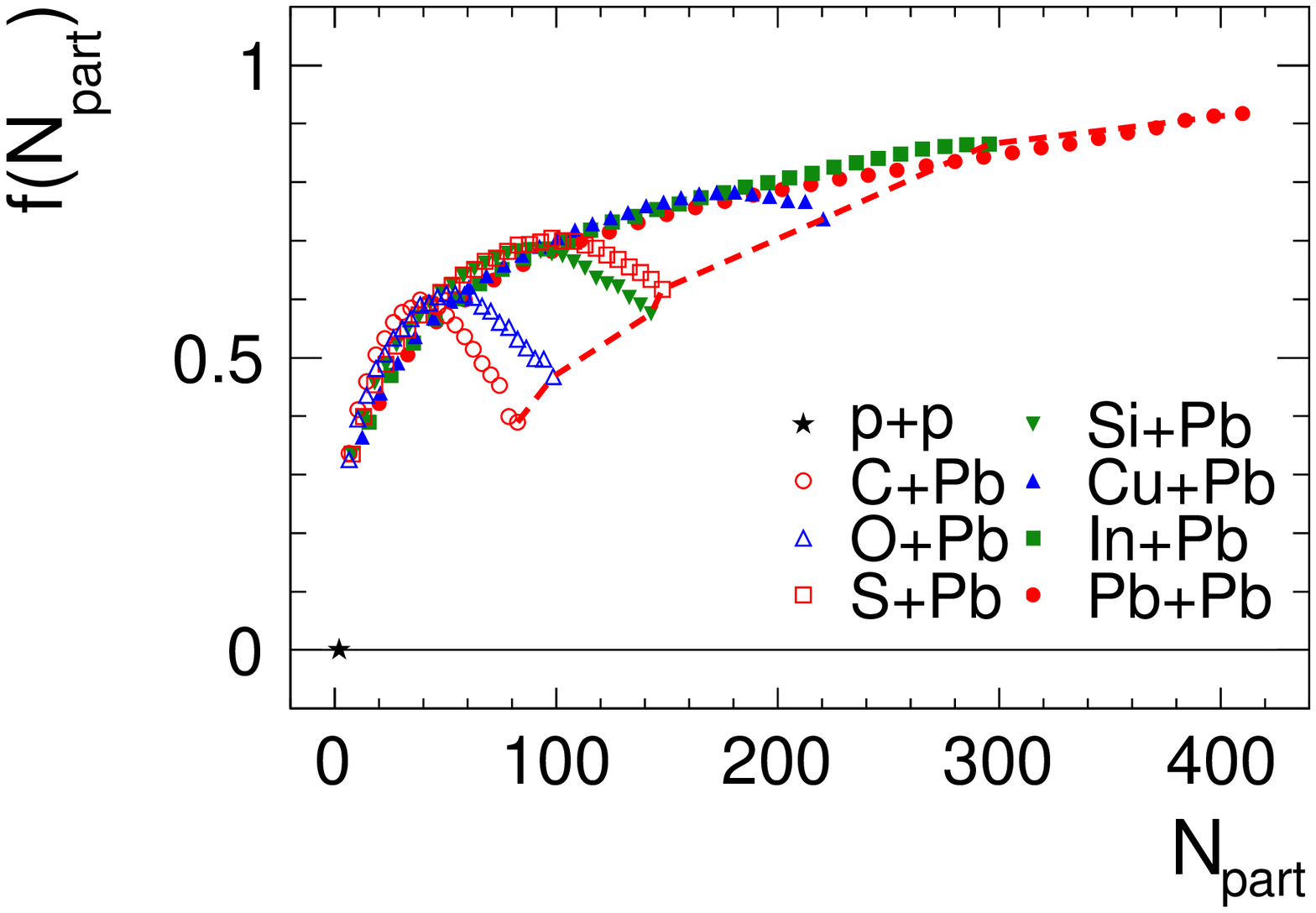}
\end{center}
\end{minipage}
\end{center}
\caption{
The fraction $f(\npart)$ of participating nucleons that scatter more than 
once as a function of the number of participants \npart.  $f(\npart)$ was
calculated with a Glauber model \cite{GLAUBER,REYGERS}.  The left panel 
shows results for symmetric systems, the right panel for asymmetric ones.
The dashed lines connect the values for the most central collisions.
}
\label{fig:glauber} 
\end{figure}
%

Another interesting aspect of the core-corona model is shown in the right
panel of \Fi{fig:glauber}.  Here, the \npart\ dependence of $f(\npart)$ is
shown for asymmetric collisions.  In this case quite distinct centrality
dependences can be observed.  While for symmetric collisions this 
dependence is following a continous rise (left panel of \Fi{fig:glauber}), 
for asymmetric collision systems with a small projectile nucleus (e.g. O+Pb 
or Si+Pb) a rapid rise followed by a maximum and a subsequent decrease of 
$f(\npart)$ is seen.  Therefore, a similar centrality dependence of particle 
yields can be expected for these type of collisions and its measurement would 
constitute a test for the validity of the core-corona model.


\section{Search for the Critical Point}
\vspace{16pt}

\subsection{Fluctuations}

\begin{figure}[t]
\begin{center}
\includegraphics[width=0.95\linewidth]{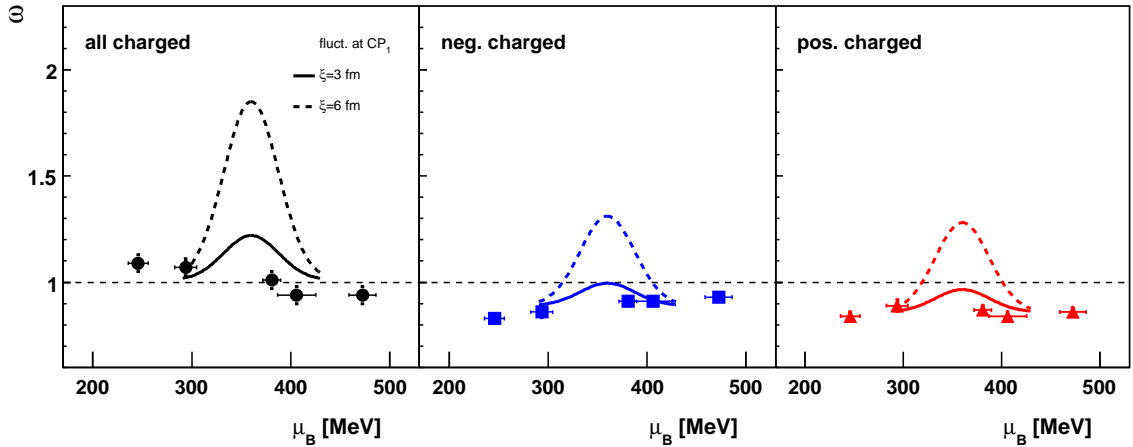}
\end{center}
\caption{
Energy dependence of multiplicity fluctuations, given by the scaled variance
$\omega$, for the 1\% most central Pb+Pb collisions in the forward rapidity
region ($1.1 < y_{\pi} < \ybeam$) as measured by the NA49 experiment 
\cite{NA49MULTFLUC,GREBIESZKOW}.  The different beam energies are represented
by the corresponding \mub~values, determined by a statistical model fit 
\cite{BECATTINI3}.  The lines correspond to predictions for for critical point 
(see text).
}
\label{fig:multfluct}
\end{figure}

One of the observables proposed to be sensitive to the presence of a
critical point in the QCD phase diagram are fluctuations of either mean
transverse momentum or multiplicity \cite{STEPHANOV1}.  If the chemical
freeze-out is happening in the vicinity of the critical point, an 
enhancement of fluctuations might be visible.  The NA49 experiment has 
therefore performed a systematic study of multiplicity fluctuations in 
very central (1\% most central) Pb+Pb collisions as a function of beam 
energy \cite{NA49MULTFLUC}.  Figure~\ref{fig:multfluct} shows the 
resulting scaled variance $\omega = \textrm{Var}(n)/\langle n \rangle$
as a function of baryonic chemical potential \mub, that has been derived
from statistical model fits to the particle abundances measured at the 
different beam energies \cite{BECATTINI3}.  

While for all charged particles taken together $\omega$ is close to
unity, it is slightly below one for negatively and positively charged
particles analyzed separately, i.e. the distributions are in these cases
a bit narrower than the corresponding Poissonian.  However, in none of 
the studied charge combinations a significant energy dependence has been
so far observed.

Figure~\ref{fig:multfluct} also shows some theoretical expectations
for $\omega$, which result from the combination of several assumptions
\cite{GREBIESZKOW}.  The position of the critical point in terms of $T$ 
and \mub\ is taken from a lattice QCD calculation \cite{FODOR1}.  The 
magnitude of the fluctuations at the critical point is based on 
\cite{STEPHANOV1,STEPHANOV2}, assuming two different correlation lengths 
$\xi = 3$~fm (solid line) and $\xi = 6$~fm (dashed line), while the widths 
of the enhancements around the critical point has been chosen according to 
\cite{HATTA1} as $\sigma(\mub) \approx 30$~MeV.  Both predictions are at 
variance with the data, so that up to now there is no evidence for a 
critical point in fluctuation measurements.

\subsection{Transverse Mass Spectra of Baryons and Anti-baryons}

Another possible way of detecting a critical point has been suggested
in \cite{ASKAWA1}.  Here it was shown that the proximity of a critical
point, or rather a critical area, might deform the isentropic trajectories 
of an expanding fireball in the $T$-$\mub$~plane, due to a focussing effect 
towards the critical point.  As a consequence one should see a decrease of
the anti-baryon/baryon ratios with increasing transverse mass, while in 
the absence of a critical point this ratio should rather be flat as a 
function of \mt.

The left panel of \Fi{fig:antib_b_spectra} shows the \pbar/p~ratio versus 
\mtmzero\ for five different beam energies \cite{NA49EDEPPR,GREBIESZKOW}.  In 
order to quantify any change in transverse mass dependence, a linear fit 
has been applied to the data.  The \lab/\lam\ and \xip/\xim~ratios have 
been analyzed in the same fashion.  The resulting energy dependence of the
slope parameters $a$ is summarized in the right panel of 
\Fi{fig:antib_b_spectra}.  No significant energy dependence can be seen
and thus this observable currently does not provide any evidence for a 
critical point.

%
%


\section{Conclusions}
\vspace{16pt}

In order to probe different areas in the QCD phase diagram, the variation
of the center-of-mass energy can be considered as the best defined control 
parameter.  Especially in the energy regime of the SPS a wide region in 
the $T$-\mub\ plane can be explored experimentally, as is evident from the 
strong dependence of the chemical freeze-out parameters on the energy here.  
The situation is much less well defined when using the system size as control
parameter.  Since the system size dependences of many observables (e.g.
strangeness production and \mtavg) seem to be well described within the
core-corona approach, it is unlikely that the freeze-out temperatures
of the fireballs vary with their size.  Rather one observes a change of
the relative admixture of the core and corona contributions.  However, the 
trajectories of the fireballs in the core might be always the same, 
irrespective of their size.  Studying centrality dependences in asymmetric 
reaction systems could provide an additional experimental test of the 
core-corona model.  In these cases this model predicts for smaller projectile 
nuclei a maximum in the relative core contribution for semi-central reactions, 
in contrast to symmetric reaction systems that reach the maximum for very 
central collisions.

Several attempts have been made to search within existing data for evidences
of a critical point, such e.g. as fluctuations and the transverse mass 
dependence of anti-baryon/baryon ratios.  So far non of these searches
yielded a positive result.  The success of future searches will on one side
mainly depend on the development of more refined and differential (i.e.
scale and $p_{\rb{t}}$~dependent) observables, that allow to distinguish 
effects due to a critical point from more trivial sources.  On the other 
side, since the magnitude of these effects might be small, experimental 
progress on reducing systematic and statistical errors will be necessary. 

%
\begin{figure}[t]
\begin{center}
\begin{minipage}[b]{0.46\linewidth}
\begin{center}
\includegraphics[width=\linewidth]{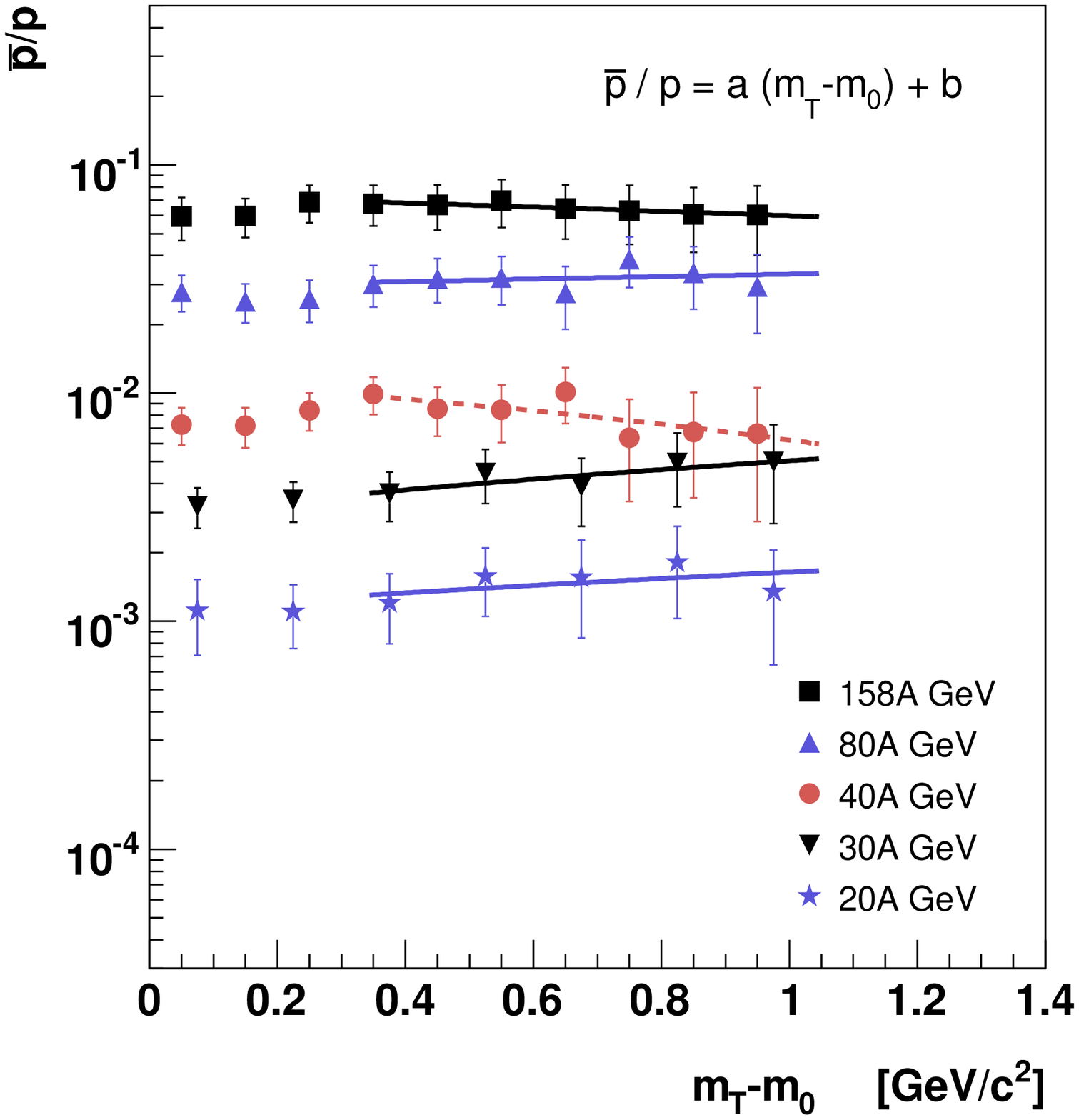}
\end{center}
\end{minipage}
\begin{minipage}[b]{0.52\linewidth}
\begin{center}
\includegraphics[width=0.95\linewidth]{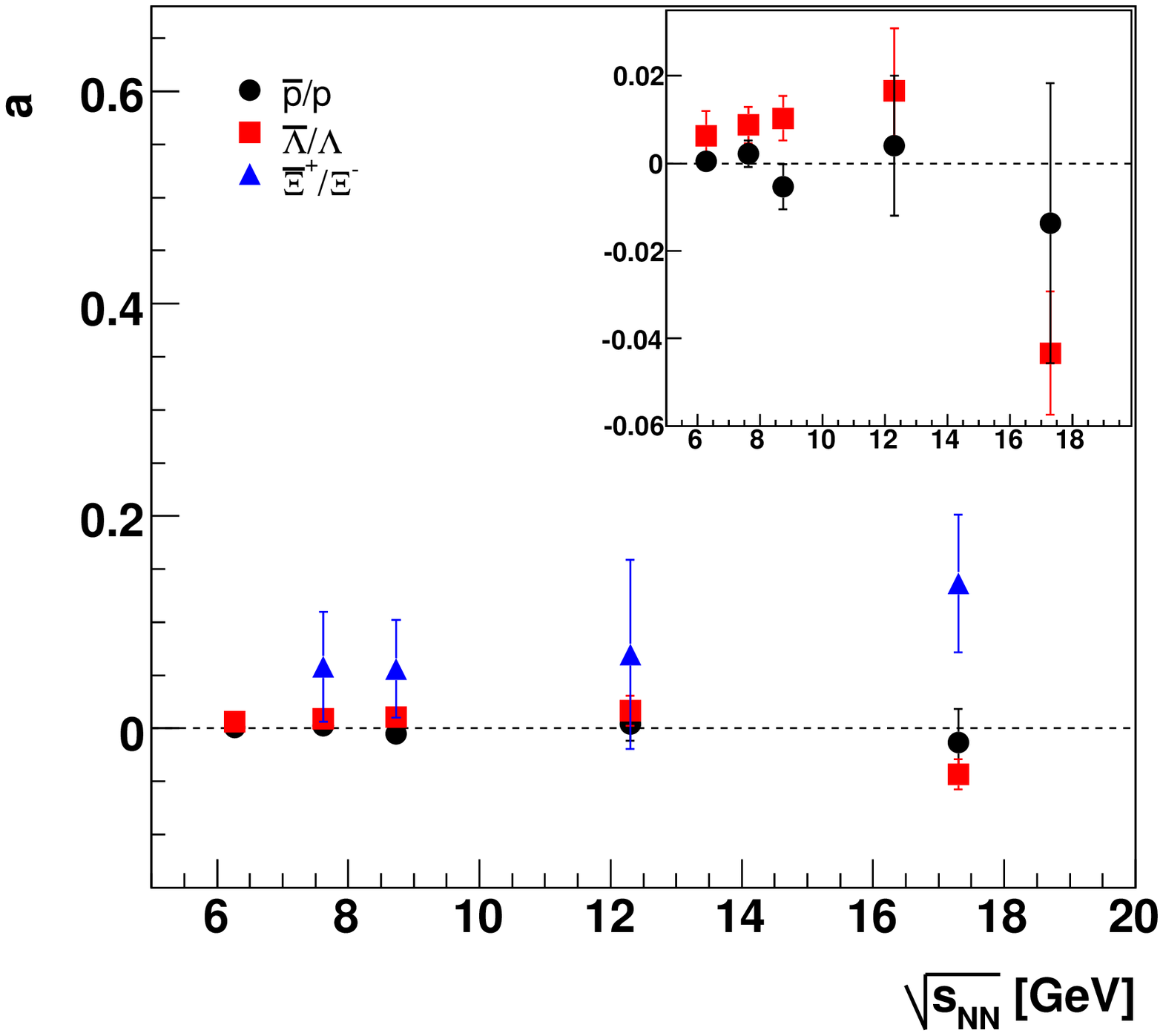}
\end{center}
\end{minipage}
\end{center}
\caption{Left: The \pbar/p~ratio versus the transverse mass \mtmzero\ 
\cite{NA49EDEPPR,GREBIESZKOW}.  The solid lines show a linear fit to 
the ratios.
Right: The energy dependence of the slope parameters $a$ from the fits
for different $\bar{\textrm{B}}$/B~ratios \cite{NA49EDEPPR,NA49EDEPHYP,
GREBIESZKOW}.  The insert shows the same data with an expanded vertical 
scale.
}
\label{fig:antib_b_spectra} 
\end{figure}
%


\section*{Acknowledgments}

The author would like to thank K.~Reygers for the help with the Glauber
model calculations and H.~Str\"{o}bele for providing helpful comments and
suggestions.



\end{document}